\documentstyle[12pt]{article}
\def\lsim{\mathrel{\rlap {\raise.5ex\hbox{$ < $}}
{\lower.5ex\hbox{$\sim$}}}}

\newcommand{\pr}{\paragraph{}}
\newcommand{\be}{\begin{equation}}
\newcommand{\ee}{\end{equation}}
\newcommand{\bea}{\begin{eqnarray}}
\newcommand{\nn}{\nonumber}
\newcommand{\eea}{\end{eqnarray}}

\newcommand{\nk}{\noindent}
\def\gappeq{\mathrel{\rlap {\raise.5ex\hbox{$>$}}
{\lower.5ex\hbox{$\sim$}}}}

\def\lappeq{\mathrel{\rlap{\raise.5ex\hbox{$<$}}
{\lower.5ex\hbox{$\sim$}}}}

\begin{document}

\begin{titlepage}
\begin{flushright}
ENSLAPP-A-508/95    \\
OUTP-95-08P   \\
\end{flushright}
\begin{centering}
\vspace{.4in}
{\large {\bf Instability of hairy black holes
in spontaneously-broken
Einstein-Yang-Mills-Higgs systems }} \\
\vspace{.4in}
{\bf E. Winstanley }    \\
\vspace{.2in}
Dept. of Physics
(Theoretical Physics), University of Oxford, 1 Keble Road,
Oxford OX1 3NP, U.K.  \\
\vspace{.1in}
and \\
\vspace{.1in}
{\bf N.E. Mavromatos$^\diamond$} \\
\vspace{.2in}
Laboratoire de Physique Th\`eorique
ENSLAPP (URA 14-36 du CNRS, associ{\'e}e \`a l' E.N.S
de Lyon, et au LAPP (IN2P3-CNRS) d'Annecy-le-Vieux),
Chemin de Bellevue, BP 110, F-74941 Annecy-le-Vieux
Cedex, France. \\
\vspace{.2in}
{\bf Abstract} \\
\vspace{.1in}
\end{centering}
{\small The stability of a new class of hairy black hole
solutions in the coupled system  of Einstein-Yang-Mills-Higgs
is examined, generalising a method suggested by
Brodbeck and Straumann and collaborators, and Volkov
and Gal'tsov. The method
maps the algebraic system of linearised radial perturbations
of the various field modes around the black hole solution
into a coupled system of radial
equations of Schr\"odinger type. No detailed knowledge of the
black hole solution is required,
except from the fact that the
boundary
conditions at the physical space-time boundaries (horizons)
must be
such so as to guarantee the {\it finiteness} of
the various expressions involved. In this way, it is demonstrated
that the above Schr\"odinger equations have bound states,
which implies the instability of the associated black hole solution.}

\vspace{1.2 in}
\pr
\nk March 1995 \\
\pr
\nk $\diamond$ On leave from P.P.A.R.C. Advanced Fellowship,
Dept. of Physics
(Theoretical Physics), University of Oxford, 1 Keble Road,
Oxford OX1 3NP, U.K.

\end{titlepage}
\newpage
Coupling gravity to non-linear systems, such as
the
non-Abelian Yang-Mills theory, or the
non-linear $\sigma$-models etc.,
has led to interesting
(classical) solutions with particle-like \cite{bartnik}
or black-hole interpretation \cite{bh}.
The interest in the latter type of solutions arises mainly
from the fact that new types of classical hair have been shown
to exist, contrary to the no-hair conjecture
characterising purely gravitational or Abelian black
holes \cite{no-hair}. This is so, because
the no-hair theorems  do not
involve the issue of stability of the solutions
in their proof, and therefore in this
respect the above classical solutions may be considered as
explicit counter-examples to these theorems.
\pr
In view of this, it is natural to
enquire into the stability of the above solutions,
which would establish their physical significance.
It has been shown that most of these systems, especially
the ones admitting particle-like interpretation, are unstable
under perturbations of the various field modes \cite{stab}.
For the black hole solutions, a corresponding general proof was
lacking so far, mainly due to the peculiar behaviour of
the stability equations on the horizons. In some cases,
however,
like the Einstein-Yang-Mills-Higgs (EYMH) systems with a
Higgs triplet, the Einstein-Skyrme (non-linear $\sigma$-model)
system, and
the Einstein-Yang-Mills-Dilaton
theory (inspired  from strings), linear stability
of the hairy solutions is established \cite{tripl}, although
non-linear stability remains an unsettled issue.
\pr
An interesting class of classical
hairy black holes has been found
recently in connection with the $SU(2)$-Einstein-Higgs
system, with a Higgs doublet as in the standard
model \cite{greene}.
These black hole solutions resemble the sphaleron solutions
in $SU(2)$ gauge theory and one would expect them to be
unstable for topological reasons.
\pr
Recently, an instability
proof of sphaleron solutions for arbitrary gauge
groups in the EYM system has been given \cite{bs,brodbeck}.
The method consists of studying linearised radial perturbations
around an equilibrium solution, whose detailed knowledge
is not necessary to establish stability.
The stability is examined by mapping the system
of algebraic equations for the perturbations
into a coupled system of differential equations
of Schr\"odinger type \cite{bs,brodbeck}.
As in the particle case of ref. \cite{bartnik}, the
instability of the solution is established once
a bound state in the respective
Schr\"odinger equations is found.
The latter shows up as an
imaginary frequency mode in the spectrum, leading to an
exponentially growing mode.
There is an elegant physical interpretation behind this
analysis, which is similar to the
Cooper pair instability of super-conductivity.
The gravitational attraction balances the
non-Abelian gauge field repulsion in the classical
solution \cite{bartnik}, but the existence of bound states
implies imaginary parts in the quantum ground state
which lead to instabilities of the solution, in much
the same way as the classical ground state
in super-conductivity is not the absolute minimum
of the free energy.
\pr
However, this method cannot be applied directly to the black
hole case, due to divergences occuring in some of the
expressions involved. This is
a result of the singular behaviour
of the metric function at the physical space-time boundaries
(horizon) of the black hole.
\pr
It is the purpose of this note to generalise the method
of ref. \cite{bs} to incorporate the black hole solution
of the EYMH system of ref. \cite{greene}. By
constructing appropriate
 trial linear radial perturbations,
following ref. \cite{brodbeck,volkov},
we show the existence of bound states in the
spectrum of the coupled Schr\"odinger
equations, and thus the instability of the black hole.
Detailed knowledge of the black hole solution is not
actually required, apart from the fact that
the existence of an horizon leads to modifications
of the trial perturbations as compared to those of
ref. \cite{bs,brodbeck},  in order to avoid divergences
in the respective expressions \cite{volkov}.
\pr
We start by sketching the basic steps \cite{bs,volkov}
that will lead to a study of the stability of
a classical solution $\phi _s (x, t)$
with finite energy
in a (generic) classical field theory.
One considers
small perturbations $\delta \phi (x,t)$
around $\phi _s  (x, t)$, and specifies \cite{bs}
the time-dependence as
\be
    \delta \phi (x ,t ) = \exp (-i \Omega t ) \Psi (x )
\label{linear}
\ee
The
linearised
system (with respect to such perturbations),
obtained from the equations of motion,
can be
cast into a Schr\"odinger eigenvalue problem
\be
   {\cal H} \Psi = \Omega ^2 A \Psi
\label{schr}
\ee
where the operators ${\cal H}$, $A$ are assumed independent
of the `frequency' $\Omega$. As we shall show later on, this is
indeed the case of our black hole solution of the EYMH system.
In that case it will also be shown that
${\cal H}$ is a self-adjoint operator with respect
to a properly defined inner (scalar) product in the space
of functions $\{\Psi \}$ \cite{bs},
and the $A$ matrix is positive definite,
$ <\Psi | A | \Psi > > 0 $.
A criterion for instability is the existence of an imaginary
frequency  mode in (\ref{schr})
\be
     \Omega ^2 < 0
\label{inst}
\ee
This is usually difficult to solve analytically
in realistic models,
and usually numerical calculations are required \cite{stab}.
A less informative method which admits analytic treatment
has been proposed recently in ref. \cite{bs,volkov},
and we shall follow this
 for the purposes of the present work.
The method consists of a variational approach
which makes use of the following functional
defined through (\ref{schr}):
\be
   \Omega ^2 (\Psi ) = \frac{ <\Psi | {\cal H } | \Psi >}
{<\Psi | A | \Psi >}
\label{funct}
\ee
with $\Psi $ a {\it trial} function.
The lowest eigenvalue is known to provide a
{\it lower} bound for this functional.
Thus,
the criterion of instability, which is equivalent to
(\ref{inst}), in this approach
reads
\bea
  \Omega ^2 ( \Psi ) &<& 0
\nn \\
    <\Psi | A | \Psi > &<& \infty
\label{trueinst}
\eea
The first of the above
conditions implies that the operator
${\cal H }$ is not positive definite, and therefore
negative eigenvalues do exist.
The second condition, on the {\it finiteness}
of the expectation value of the operator $A$,
is required to ensure that $\Psi$ lies in the
Hilbert space containing the domain of ${\cal H}$.
 In certain cases, especially
in the black hole case,
there are divergences
due to singular behaviour of modes at, say, the horizons,
which could spoil these conditions
(\ref{trueinst}).
The advantage of the above variational method
lies in the fact that it is an easier task
to choose appropriate trial functions $\Psi $
that satisfy (\ref{trueinst}) than solving the
original eigenvalue problem (\ref{schr}).
In what follows we shall apply this second method
to the black hole solution of ref. \cite{greene}.
\pr
We start by reviewing the basic formulas
for a study of stability
issues
of spherically symmetric black hole solutions
of the EYMH system \cite{greene}.
The space-time metric
takes the form \cite{greene}
\be
ds^2 =-N(t,r) S^{2}(t,r) dt^2 + N^{-1} dr^2 +
r^2 (d\theta ^2 + \sin^2 \theta d\phi ^2)
\label{one}
\ee
and we assume the following ansatz for
the non-abelian gauge potential \cite{greene,bs}
\be
A  = a_0 \tau _r dt + a_1 \tau _r dr + (\omega -1 )
[ \tau _\phi d\theta - \tau _\theta \sin \theta d\phi ]
+ {\tilde \omega} [ \tau _\theta d\theta + \tau _\phi \sin \theta d\phi ]
\label{two}
\ee
where $\omega, {\tilde \omega}$ and $a_i, i = 0,1 $
are functions of $t, r$. The $\tau _i$ are appropriately normalised
spherical
generators of the SU(2) group in the notation
of ref. \cite{bs}.
\pr
The Higgs doublet assumes the form
\be
\Phi \equiv \frac{1}{\sqrt{2}} \left( \begin{array}{c}
\nonumber  \psi _2  + i \psi _1 \\
           \phi - i \psi _3  \end{array}\right)
\qquad ; \qquad {\mbox {\boldmath $ \psi $}}
 = \psi {\mbox {\boldmath ${\hat r}$}}
\label{three}
\ee
with the Higgs potential
\be
   V(\Phi )=\frac{\lambda }{4} ( \Phi ^{\dag}  \Phi  - v^2)^2
\label{four}
\ee
where $v$ denotes the v.e.v. of $\Phi $ in the non-trivial
vacuum.
\pr
The quantities $\omega, \phi$ satisfy the static field
equations
\bea
  N \omega '' &+& \frac{(N S)'}{S} \omega ' = \frac{1}{r^2}
(\omega ^2 - 1) \omega + \frac{\phi ^2}{4}
(\omega - 1)    \nn \\
N \phi '' &+& \frac{(N S)'}{ S} \phi ' + \frac{2N}{r} \phi '
= \frac{1}{2r^2} \phi (\omega -1 )^2 + \lambda \phi
(\phi ^2 - v^2 )
\label{five}
\eea
where the  prime denotes differentiation with respect
to $r$. For later use, we also mention that
a dot
will denote
differentiation
with respect to $t$.
\pr
If we choose a gauge in which $\delta a_{0} =0$,
the linearised perturbation equations decouple into two sectors
\cite{bs} . The first consists of the gravitational modes
$\delta N$, $\delta S$, $\delta \omega$ and
$\delta \phi$ and the second of the matter perturbations
$\delta a_{1}$, $\delta {\tilde {\omega }}$
and $\delta \psi $.
In our analysis it will be sufficient
to concentrate on the matter perturbations, setting the
gravitational perturbations  $\delta N$ and $\delta S$
 to zero, because an instability will show up in
this sector of the theory.
The equations for
the linearised matter perturbations
take the form \cite{bs}
\be
   {\cal H} \Psi + A {\ddot \Psi } = 0
\label{six}
\ee
with,
\be
\Psi = \left( \begin{array}{c}
\nonumber  \delta a_1  \\
\nonumber  \delta {\tilde \omega}  \\
\nonumber  \delta \psi    \end{array}\right)
\label{seven}
\ee
and,
\be
A = \left( \begin{array}{ccc}
  Nr^2 &   0  & 0 \\
  0 &  2   & 0 \\
  0 &  0   & r^2  \\
\end{array}\right)
\label{eight}
\ee
and the components of ${\cal H}$ are
\bea
{\cal H}_{a_1a_1} &=& 2 (N S)^2 \left( \omega ^2
+ \frac{r^2}{8} \phi ^2 \right) \nn \\
{\cal H}_{{\tilde \omega} {\tilde \omega}} &=&
2 p_* ^2 + 2NS^2 \left( \frac{\omega ^2 -1}{r^2} + \frac{\phi ^2}{4}
\right)
\nn \\
{\cal H}_{\psi\psi} &=& 2 p_*\frac{r^2}{2} p_* +
2 NS^2 \left( \frac{(\omega + 1)^2}{4} + \frac{r^2 }{2}\lambda
(\phi ^2 - v^2) \right) \nn \\
{\cal H}_{a_1{\tilde \omega}} &=& 2i N S [ (p_* \omega) - \omega p_* ]
\nn \\
{\cal H}_{{\tilde \omega} a_1} &=&
2i [ p_* N S \omega + N S (p_* \omega) ]
\\
{\cal H}_{a_1 \psi } &=& \frac{i r^2}{2} N S [(p_* \phi) - \phi p_* ]
\nn \\
{\cal H}_{\psi a_1} &=& i p_* \frac{r^2}{2}
NS \phi + i\frac{r^2}{2} NS (p_* \phi )
\nn \\
{\cal H}_{{\tilde \omega}\psi} &=& {\cal H}_{\psi {\tilde
\omega}} = -\phi N S^2     \nn
\label{nine}
\eea
where the operator $p_*$ is
\be
    p_* \equiv - i NS \frac{d}{dr}
\label{ten}
\ee
Upon specifying the time-dependence
(\ref{linear})
\be
  \Psi (r, t) = \Psi (r) e^{i \Omega t}  \qquad ;
\qquad
\Psi (r) = \left( \begin{array}{c}
\nonumber  \delta a_1 (r) \\
\nonumber  \delta {\tilde \omega} (r) \\
\nonumber  \delta \psi (r)   \end{array}\right)
\label{linar2}
\ee
one arrives easily to an eigenvalue
problem of the form (\ref{schr}), which can then be
extended to the variational approach (\ref{trueinst}).
\pr
To this end, we choose
as trial perturbations the following expressions
(c.f. \cite{bs})
\bea
\nonumber  \delta a_1 &=& \omega ' Z  \\
\nonumber  \delta {\tilde \omega} &=& (\omega ^2 - 1) Z  \\
\nonumber  \delta \psi &=& \frac{1}{2} \phi (\omega - 1) Z
\label{eleven}
\eea
where $Z$ is a function of $r$ to be determined.
\pr
One may define the inner product
\be
  <\Psi | \Phi > \equiv \int _{r_h} ^{\infty} {\overline \Psi } \Phi
\frac{1}{NS} dr
\label{twelve}
\ee
where $r_h$ is the position of the horizon of the black hole.
The operator ${\cal H}$ is then symmetric with respect to
this scalar product.
Following ref. \cite{bs}, consider the expectation value
\be
  <\Psi | A | \Psi > = \int _{r_h}^{\infty}
dr \frac{1}{NS} Z^2 \left[ Nr^2 (\omega ')^2 +
2 (\omega ^2 - 1)^2 +
\frac{r^2}{4} \phi ^2 (\omega - 1)^2 \right]
\label{thirteen}
\ee
which is clearly positive definite for real $Z$.
Its finiteness will be examined later, and depends
on the choice of the function $Z$.
\pr
Next, we proceed to the evaluation of
the expectation value of the
Hamiltonian ${\cal H}$ (\ref{nine}); after a tedious calculation
one obtains
\bea
  <\Psi | {\cal H} | \Psi > &=& \int _{r_h}^{\infty}
dr S Z^2 \{ - 2 N (\omega ')^2 + 2 P^2 N (\omega ^2 - 1)^2
 \nn \\  &  &
+ \frac{1}{4} P^2 N r^2 \phi ^2 (\omega - 1)^2
- \frac{2}{r^2} (\omega ^2 - 1)^2  - \frac{1}{2}
\phi ^2 (\omega - 1)^2 \}  \\
\nn  & + & {\mbox {boundary terms}}
\label{fourteen}
\eea
where $P \equiv \frac{1}{Z} \frac{d Z}{d r} $.
The boundary terms will be shown to vanish so we omit them in
the expression (\ref{fourteen}). The final result is
\bea
<\Psi | {\cal H} | \Psi >  &= & \int _{r_h}^{\infty}
dr S \left\{ - 2 N (\omega ')^2 - \frac{2}{r^2}
(\omega ^2 - 1)^2 - \frac{1}{2} \phi ^2 ( \omega - 1)^2   \right\}
 \nn \\
 & + & \int _{r_h}^{\infty} dr \left\{
\frac{2}{r^2} (\omega ^2 - 1) ^2 + \phi ^2 (\omega - 1)^2
+ 2 N (\omega ')^2 \right\} S ( 1 - Z^2)  \nn \\
 & +  & \int _{r_h}^{\infty} dr S N \left( \frac{d Z}{dr} \right) ^2
\left[ 2 (\omega ^2 -1)^2 + \frac{1}{4} r^2 \phi ^2 (\omega - 1)^2
\right]
\label{fifteen}
\eea
\pr
The first of these terms is manifestly negative.
To examine the remaining two, we introduce the
`tortoise' co-ordinate $r^*$ defined by \cite{volkov}
\be
      \frac{d r^*}{dr} = \frac{1}{N S}
\label{sixteen}
\ee
 and define a sequence of functions $Z _k ( r^* )$ by
\cite{volkov}
\be
  Z_k ( r^* ) = Z\left( \frac{r^*}{k}
\right) \qquad ; \qquad k =1,2, \ldots
\label{seventeen}
\ee
where
\bea
Z ( r^* ) &=& Z ( -r^* ), \nn \\
Z ( r^* ) &=& 1 \qquad
{\mbox {for $ r^* \in [ 0, a] $}}  \nn \\
 - D \le&  \frac{d Z}{d r^*} & < 0, \qquad
{\mbox { for $r^* \in [a, a + 1 ]$}}
\nn \\
 Z( r^* ) &=& 0  \qquad
{\mbox {for $ r^* > a + 1 $}}
\label{eighteen}
\eea
where $a$, $D$ are arbitrary positive constants.
Then, for each value of $k$ the vacuum expectation values
of ${\cal H }$ and $A$ are finite,
$ <\Psi | {\cal H } | \Psi > < \infty $,
and $<\Psi |A| \Psi > < \infty $,
with $Z = Z_k $, and all boundary terms vanish. This
justifies {\it a posteriori} their being dropped in
eq. (\ref{fourteen}).  The integrands in the second and third terms
of eq. (\ref{fifteen}) are uniformly convergent
and tend to zero as $ k \rightarrow \infty $. Hence, choosing
$k $ sufficiently large the dominant contribution
in (\ref{fifteen}) comes from the first term which is negative.
\pr
This confirms the existence of bound states in the
Schr\"odinger equation (\ref{six}), (\ref{schr}),
and thereby the instability (\ref{trueinst})
of the associated black hole solution of ref. \cite{greene}
in the coupled EYMH system.
\pr
The above analysis reveals the existence of at least one
negative {\it odd-parity} eigenmode
in the spectrum
of the
EYMH black hole, which implies its instability.
The exact number of such negative modes is an interesting
question and we plan to investigate it in the near future.
Recently, a method for determining the number of the
sphaleron-like unstable modes
has been applied
by Volkov et al. \cite{eigenmodes}
to the gravitating sphaleron case, and one might
be able to extend it to the present EYMH black hole.
According to the analysis of ref. \cite{eigenmodes},
for EYM black holes, there are $n$ - unstable sphaleron-like
modes under radial perturbations, where $n$ is the number of
nodes
of the equilibrium solution \cite{bartnik}.
This number does not depend on the details of the equilibrium
solution, such as the horizon
geometry, size etc. This is due to the topological nature of the
instabilities. In this respect,
we mention that
an interesting
connection could be made
with the global analysis
of ref. \cite{maeda}, where catastrophe theory
was invoked to provide a way of evaluating
the number of unstable modes of certain (non-sphaleron)
black hole
solutions.
{}From the global analysis of ref. \cite{maeda}
there are other non-sphaleronic
types of non-Abelian black holes, whose
`high entropy' phase is stable. In our analysis, this would imply
an extension of the variational approach to incorporate
finite temperature effects for the matter perturbations
in non-sphaleron black holes\footnote{For sphaleron-like
black holes, the topological nature of the instability
might complicate
the connection with catastrophe theory if the number of unstable
modes is independent of the details of the equilibrium
solution, as appears to be the case for the EYM
system \cite{eigenmodes}.}.
The finite temperature would be a result of
the existence of the
horizon entropy associated with the black hole in a semi-classical
analysis.
It might well be that the number of unstable modes
of these (non-sphaleron) black holes is somehow
affected by the temperature, in the sense that above
a `critical' temperature (corresponding to a certain
horizon size) the
bound states of the Schr\"odinger
equation (\ref{schr}) disappear, or their number is reduced. This
would correspond
to the high-entropy `stable' black holes of ref. \cite{maeda},
in the sense of the catastrophe theory.
At present, such issues remain open. We hope to come back to
these in the near future.
\pr
\pr
\nk {\Large {\bf Acknowledgements } }
\pr
We thank
K. Tamvakis and P. Kanti for discussions.
One of us (E.W.) would like to thank CERN, Theory Division,
for the hospitality during the initial stages of this work.
She also thanks E.P.S.R.C. (U.K.) for a research
studentship.
The work of N.E.M. is supported by a EC Research Fellowship,
Proposal Nr. ERB4001GT922259.

\end{document}